\documentstyle[12pt]{article}
\begin{document}
\title{Complexified quantum field theory in operator form}
\author{Slobodan Prvanovi\'c \\
{\it Institute of Physics Belgrade, } \\
{\it University of Belgrade, Pregrevica 118, 11080 Belgrade,}\\
{\it Serbia}}
\date{}
\maketitle

\begin{abstract}
The anti self-adjoint operators of imaginary coordinate and momentum, together with the self-adjoint operators of real coordinate, momentum, energy and time are 
used in construction of the quantum field theory in operator form. This formalism, being free of many dubious mathematical situations characteristic to the 
standard treatment of quantum fields, is applied in formulation of the new gauge condition for electromagnetic field that describes the absence of the scalar and 
longitudinal photons. Proposed formalism offers adequate theoretical framework for proper description of the multicomponent fields, including quantum gravity.   

Keywords: anti self-adjoint operator; negative energy; Hamiltonian; gauge condition; quantum gravity
\end{abstract}

\section{Introduction}

Quantum field theory (QFT) is a century old theory and still attracts much attention from variety of reasons [1-21]. Its development was not straightforward and even 
now it can not be said that we have complete theory of all known quantum fields. Importance of this theory, we believe, lies in the fact that it should be the 
fundamental theory of Nature. Quantum mechanics, on the other side, should appear as the special case of the appropriate field theory for a single or countably many 
particles. 

QFT was overwhelmingly successful during last few decades in a sense that it offered comprehensive and accurate theoretical description of many experiments, 
especially within particle physics and condensed matter physics. But, on the other side, it is not free of problems, unsatisfactory solutions and misconceptions 
[22-24]. All of these, we believe, emerge from the inappropriate mathematical foundation of QFT, or in Robbert Dijkgraafs words: "Quantum field theory emerged as an 
almost universal language of physical phenomena, but it's in bad math shape" [25]. 

In this article, we want to propose new formalism that will respect all good features of QFT and avoid some theoretical obstacles. More concretely, It is well 
known that standard formalism of QFT uses creation and annihilation operators and that these operators are accompanied with functions $e^{\pm{1\over {i \hbar}} 
 \vec k \cdot \vec r}$. This inconsistency, {\it i. e.}, the mix of abstract operators and explicit functions, is the primary source of problems in 
QFT. Beside these functions, there are $e^{\pm {1\over {i \hbar}} E \cdot t}$ which lead to appearance of negative energies that, within the standard formulation of 
QFT (as well as within the standard formulation of quantum mechanics), can not be explained in satisfactory manner. In this respect, we are going to propose new 
formalism of QFT that is completely in the operator form. The above mentioned functions will appear only after appropriate representation of the involved operators. 
Since we are going to introduce operators of energy and time, this will hold for both $e^{\pm {1\over {i \hbar}} \vec k \cdot \vec r}$ and 
$e^{\pm {1\over {i \hbar}} E \cdot t}$. Moreover, we will introduce anti self-adjoint operators of coordinate and momenta which will allow us to treat negative 
energies in a consistent way.    

The formalism we are going to introduce will allow us to avoid states that have negative norm. The experimental fact that singular and longitudinal photons 
have not been found will be connected to the new gauge condition that is going to be introduced. Moreover, new formalism offers possibility to treat gravitation on 
an equal footing with other fields regarding its quantum formulation. In a forthcoming article we shall discuss spinor fields within the framework of this formalism, 
together with the representations of simmetries, and that will allow straightforward introduction of the interaction terms in the Hamiltonian of interacting fields.    

Before introducing new formalism, which will be done in Section 4., we shall shortly review our proposal of the operators of time and energy in Section 2. 
and anti self-adjoint operators of coordinate and momenta in Section 3. Some remarks will be given in the last section. 
 
\section{Operators of time and energy}

In [26-32] we have proposed the formalism of the operators of time and energy. There are many articles where the operator of time is discussed, {\it e.g.} [33-35] 
and references therein, and our approach is similar to [36], and references therein, and [37]. The starting point of our approach is to treat time and energy equally 
with coordinate and momentum. From this, it follows that the separate Hilbert space ${\cal H}_t$, where the operators of time $\hat t$ and energy 
$\hat s$ should act, has to be introduced and that should be done in the same manner as it is done for each degree of freedom in the standard formulation of quantum 
mechanics. Then, the commutation relation for these operators is imposed in the same way as it is done for the operators of coordinate and momentum:
\begin{equation}
{1\over i\hbar} [\hat t , \hat s ] = - \hat I .
\end{equation}
This leads to the unbounded spectrum of these operators $( -\infty , +\infty )$, just like it is the case for the operators of coordinate and momentum. The 
eigenvectors of $\hat t$ are $\vert t \rangle $ for every $t\in {\bf R}$. In time $\vert t \rangle$ representation, the operator of energy is differential operator, 
just like the operator of momentum is in coordinate representation, and it is given by $i \hbar {\partial \over \partial t}$. Its eigenvectors $\vert E \rangle$ in 
this representation are $e^{{1\over i\hbar} E\cdot t}$ for every $E\in \bf R$. 

So, within this formalism there are negative energies (in similarity with the negative values of momentum in the standard quantum mechanics). But, the Schr\"odinger 
equation is a constraint in the overall Hilbert space that selects the states with non-negative energy for the usually used Hamiltonians. More precisely, for the 
quantum system with three spatial degrees of freedom the complete space is ${\cal H}_x \otimes {\cal H}_y \otimes {\cal H}_z \otimes {\cal H}_t$. The operators of 
coordinate are $\hat r _x$, $\hat r _y$ and $\hat r _z$ and operators of momentum are $\hat k _x$, $\hat k _y$ and $\hat k _z$ (acting non trivially in appropriate 
spaces and being identical to those used in the standard formalism of quantum mechanics). The Hamiltonian is $H ({\hat {\vec r}}, {\hat {\vec k}} )$ 
and the constraining equation:
\begin{equation}
 \hat s \vert \psi \rangle  = H ({\hat {\vec r}}, {\hat {\vec k}} ) \vert \psi \rangle  ,
\end{equation}
is nothing else but the Schr\"odinger equation. Namely, by taking the coordinate-time $\vert x \rangle \otimes \vert y \rangle \otimes \vert z \rangle \otimes 
\vert t \rangle $ representation of the previous equation, one gets the familiar form of the Schr\"odinger equation:
\begin{equation}
 i \hbar {\partial \over \partial t } \psi (x, y, z, t)  = H ({\vec r} , - i \hbar {\partial \over \partial {\vec r} } ) \psi (x, y, z, t)  .
\end{equation}
So, the states with non-negative energies, that are only considered within the standard formalism of quantum mechanics, are selected by the Schr\"odinger equation 
due to the non-negative spectra of the Hamiltonian $H ({\hat {\vec r}}, {\hat {\vec k}} )$. However, as is well known, the negative energies do appear 
in QFT, so the negative eigenvalues of $\hat s$ will be discussed below.

\section{Anti self-adjoint operators and negative energies}

The standard quantum mechanics, with its self-adjoint operators of coordinate and momentum and appropriate real eigenvalues, is a part of the complete theory that 
involves anti self-adjoint operators of coordinate and momentum and their imaginary eigenvalues, as well. The last ones, together with the negative energies, find 
their natural explanation within the complexified quantum mechanics and let us give here a short recapitulation of this subject that we have discussed in [38]. 

Beside the self-adjoint operators of coordinate and momentum ${\hat {\vec r}} _{re} $ and ${\hat {\vec k}} _{re}$, acting in the rigged Hilbert spaces 
${\cal H}_{x, re} \otimes {\cal H}_{y, re} \otimes {\cal H}_{z, re} $, (where the index $re$ stands for the real spectrum of these operators) one can 
introduce the anti self-adjoint operators of coordinate and momentum ${\hat {\vec r}} _{im}$ and ${\hat {\vec k}} _{im}$, acting in the rigged Hilbert 
spaces ${\cal H}_{x, im} \otimes {\cal H}_{y, im} \otimes {\cal H}_{z, im} $, (where the index $im$ stands for the imaginary spectrum of these operators). The spectral 
form of the imaginary coordinate $\hat r_{x,im}$, in the basis $\vert r_{x, im} \rangle$, where $x_{im}$ ranges over entire imaginary axis, is $\hat r_{x,im} = 
\int r_{x,im} \vert r_{x,im} \rangle \langle r_{x,im} \vert dr_{x,im}$, where $dr_{x,im}$ is the real measure (similarly holds for other two imaginary coordinates 
$\hat r_{y,im}$ and $\hat r_{z,im}$). In the basis $\vert r_{x,im} \rangle$, the anti self-adjoint operator of imaginary momentum $\hat k_{x,im}$ is represented by 
$-i \hbar {\partial \over \partial r_{x,im}}$. Its eigenvectors, in the same representation, are the "imaginary" plain waves $e^{-{1\over i \hbar} k_{x,im} \cdot 
r_{x,im}}$ with the imaginary eigenvalues. In parallel to the case of the real coordinate and momentum, the commutator of the imaginary ones is proportional to the 
$\hat I _{im}$, which is $\hat I _{im} = \int \vert r_{x,im} \rangle \langle r_{x,im} \vert dr_{x,im} $. (Similarly holds for other two imaginary momenta 
$\hat k_{y,im}$ and $\hat k_{z,im}$.)

If one uses Hamiltonian that is, for example, quadratic in imaginary momentum, then its spectrum would be $(- \infty , 0]$. This means that the negative energies 
are connected to the imaginary coordinate and momentum and for these anti self-adjoint operators they are as natural as the positive energies are for the real 
coordinate and momentum. On the other side, there are situations (like tunneling [38]) where one can find out that the imaginary values of momentum are unavoidable. 
This indicates that we should accept the fact that there are two worlds - one that is characterized by real numbers (our world) and one that is characterized by 
imaginary numbers. These two worlds together, not just the world that we know very well, constitute the whole universe. The complete formalism has to contain both the 
real and the imaginary spatial coordinates and momentum. Both positive and negative values of energy should be treated as possible and the particular choice of the 
Hamiltonian would select, via Schr\"odinger equation, which one would be realised by the quantum systems state. 

The negative energies that are always present in the standard formulation of QFT indicate that the inappropriate formalism is used. These energies can not be explained 
if only real coordinates and momenta are considered. It is necessary to enlarge the algebra of operators with the imaginary ones and this is what we are going to do 
in the sequel.         

\section{Quantum fields in operator form}

Within the standard formalism of QFT, the single component quantum field is:
\begin{equation}
\hat \phi (\vec r , t) \sim
\int ( \hat a_{\vec k} e^{-{1\over {i \hbar}} \vec k \cdot \vec r }  e^{+{1\over {i \hbar}} E_{\vec k} \cdot t } + 
\hat a^{\dagger} _{\vec k} e^{+{1\over {i \hbar}} \vec k \cdot \vec r } e^{-{1\over {i \hbar}} E_{\vec k} \cdot t }) \cdot 
d\vec k .
\end{equation}
Obviously, if for $ e^{+{1\over {i \hbar}} E_{k} \cdot t}$ one finds positive energy, then for $  e^{-{1\over {i \hbar}} E_{k} \cdot t }$ there 
would be negative energy (or frequency). This is in dissonance with the rest of the formalism, {\it i. e.}, this negative energy term can not be 
explained within the standard formalism of QFT. Then, if one is analyzing situations characterized with the multiplication of fields, for example 
when one is concerned about commutation of the field with itself, {\it i. e.}, about microcausality, then there emerge the problem with the term 
$ e^{+{1\over {i \hbar}} \vec k \cdot \vec r } e^{-{1\over {i \hbar}} \vec k \cdot \vec r \ ' }$.   

In order to address these, as well as other issues, we propose formalism which is based on the operators. Let us introduce this formalism heuristically.

The functions $e^{\pm {1\over {i \hbar}} \vec k \cdot \vec r }$ representing the modes of the field are seen as the coordinate representation 
of the vector $\vert \vec k \rangle = \vert k_x \rangle \otimes \vert k_y \rangle \otimes \vert k_z \rangle $. Then, since $ e^{\pm {1\over {i \hbar}} 
\vec k \cdot \vec r }$ stand by the operators of creation and annihilation, we shall not use vectors $\vert \vec k \rangle$, but the 
operators - dyads, $\vert \vec k \rangle \langle \vec k \vert$. Since the time and spatial coordinates should be treated equally, 
instead of $ e^{\pm { 1\over {i \hbar}} E_{\vec k} \cdot t}$, that are seen as the time representation of the energy states, we shall use $\vert \pm E 
\rangle \langle \pm E \vert$. Independence of degrees of freedom should be represented by introduction of separate Hilbert spaces which should be directly multiplied. 

The proper explanation of the appearance of negative energies in QFT demands introduction of the imaginary momentum and the sectors of the formalism that are attached 
to the real world and "imaginary world" should be independent and formally the same. Therefore, the mode of the quantum field characterized by 
${\hat {\vec k}} _{re}$, $E _{\vec k_{re}}$, ${\hat {\vec k}} _{im}$ and $E _{\vec k_{im}}$ is represented with:
\begin{equation}
\vert \vec k_{re} \rangle \langle \vec k_{re} \vert \otimes \vert E_{\vec k_{re}} \rangle \langle E_{\vec k_{re}} 
\vert \otimes 
\vert \vec k_{im} \rangle \langle \vec k_{im} \vert \otimes \vert E_{\vec k_{im}} \rangle \langle E_{\vec k_{im}} \vert .
\end{equation}
The values of $E _{\vec k_{re}}$ and $E _{\vec k_{im}}$ depend on the Hamiltonian, which will become clear below. Since both 
$E _{\vec k_{re}}$ and $E _{\vec k_{im}}$ are real, one can take one rigged Hilbert space where $  \vert E_{\vec k_{re} , 
\vec k_{im}} \rangle \langle E_{\vec k_{re} , \vec k_{im}} \vert $ would represent the energy of the mode. However, we shall take 
two spaces and separate energy in a part that is connected to the real world and the one that is connected to the imaginary world.  

In the rigged Hilbert spaces where above dyads appear, the operators of spatial coordinates ${\hat {\vec r}} _{re}$ and ${\hat {\vec r}} _{im}$ and 
momenta ${\hat {\vec k}} _{re}$ and ${\hat {\vec k}} _{im}$ act. Beside these, as the components of the quadri vectors of real and imaginary 
coordinates and momenta, there are operators of energy and time, as well. For example, $\hat r ^0 _{im} = c_{im} \cdot \hat t$ and $\hat k ^0 _{im} = 
{\hat s \over c_{im}}$, where $c_{im}$ is the speed of light in the imaginary world, which is $c_{im} = i \cdot c $, ($c_{re} = c$). 

Since the fields constitute, so to say, of the normal modes and the amplitude, the additional rigged Hilbert is needed. Within this space, the standard 
non-commuting operators $\hat q$ and $\hat p$ act. With these, one can introduce:
\begin{equation}
\hat q \otimes 
\vert \vec k_{re} \rangle \langle \vec k_{re} \vert \otimes \vert E_{\vec k_{re}} \rangle \langle E_{\vec k_{re}} 
\vert \otimes \vert \vec k_{im} \rangle \langle \vec k_{im} \vert \otimes \vert E_{\vec k_{im}} \rangle \langle 
E_{\vec k_{im}} \vert ,
\end{equation}
\begin{equation}
\hat p \otimes  
\vert \vec k_{re} \rangle \langle \vec k_{re} \vert \otimes \vert E_{\vec k_{re}} \rangle \langle E_{\vec k_{re}} 
\vert \otimes \vert \vec k_{im} \rangle \langle \vec k_{im} \vert \otimes \vert E_{\vec k_{im}} \rangle \langle 
E_{\vec k_{im}} \vert .
\end{equation}
Now, one can define the single component quantum field in the operator form as:
\begin{equation}
\hat \phi = \int \int \hat n \otimes  
\vert \vec k_{re} \rangle \langle \vec k_{re} \vert \otimes \vert E_{\vec k_{re}} \rangle \langle E_{\vec k_{re}} 
\vert \otimes \vert \vec k_{im} \rangle \langle \vec k_{im} \vert \otimes \vert E_{\vec k_{im}} \rangle \langle 
E_{\vec k_{im}} \vert \cdot d \vec k_{re} \cdot d \vec k_{im} .
\end{equation}  
It is understood that the number operator $\hat n = \hat a^{\dagger} \cdot \hat a$ and that the operators of creation and annihilation are 
constructed from $\hat q$ and $\hat p$ in the familiar way. Needless to say, the "amplitude" operators $\hat q$ and $\hat p$, being formally the same as the 
corresponding spatial ones, {\it e. g.} $\hat r_x$ and $\hat k_x$, non-trivially act in their own space, while there are eight directly multiplied rigged Hilbert 
spaces where operators of spatial, both real and imaginary, coordinates and momenta, time and energy act. (Each of these operators non-trivially acts only in 
one space, so they actually are, for example, $\hat I \otimes \hat p_x \otimes \hat I \otimes \hat I \otimes \hat I \otimes \hat I \otimes \hat I \otimes \hat I 
\otimes \hat I$.)
      
The Hamiltonian is constructed from the spatial operators, for example:
\begin{equation}
\hat H = +({\hat {\vec k}} _{re} ^2 \cdot c_{re} ^2 + m^2  \cdot c_{re} ^4 )^{1\over 2} - 
({\hat {\vec k}} _{im} ^2 \cdot c_{im} ^2 + m^2 \cdot c_{im} ^4 )^{1\over 2} .
\end{equation}
For $m=0$ this is the Hamiltonian of electromagnetic field, while for non-vanishing mass it is related to the Proca field. The consistency of the modes, and entire 
field, follows from the Schr\"odinger like dispersion equation:
\begin{equation}
\hat H \cdot \hat \phi = \hat S \cdot \hat \phi ,
\end{equation}
where $\hat S = \hat s _{re} + \hat s _{im}$, $\hat s _{re}$ and $\hat s _{im}$ act on $\vert E_{\vec k_{re}} \rangle$ and 
$\vert E_{\vec k_{im}} \rangle$, 
respectively (indices of these operators of energy designate the space where they act, and not their spectrum which is real). 

The state of the field is:
\begin{equation}
\sum c_{\vec k _{re} , \vec k _{im}} \vert n_{\vec k _{re} , \vec k _{im}} \rangle \otimes \vert \vec k _{re} 
\rangle \otimes \vert E _{\vec k _{re}} \rangle \otimes \vert \vec k _{im} \rangle \otimes \vert E _{\vec k _{im}} \rangle .
\end{equation}
Within the standard treatment of quantum fields, in expressions for the fields the value of momentum in $e^{-{1\over {i \hbar}} \vec k \cdot 
\vec r }$ and $e^{+{1\over {i \hbar}} \vec k \cdot \vec r }$ is the same. If one takes for $\vec k _{re}$ and 
$\vec k _{im}$ in (11) to be $\vec k _{re} = \vec k$ and $\vec k _{im} = i \cdot \vec k$, then for the 
Hamiltonian (9), the overall energy would be equal to zero and there would be no problems with the infinite energy of the fields. 

In general, there could be many fields with many components. From the mathematical point of view, all of them should be treated equally. On the other hand, 
physics dictates some constraining equations that entangles them and which should be satisfied when the involved operators act on states. By constraining equations 
we mean dispersion, gauge and diffeomorphism equations for example, and to the states for which all imposed equations are satisfied we may attach the physical 
meaning.  

If a multicomponent field should be formalised, then for each of the components there should be in direct product separate "amplitude" space and appropriate 
operators, while the "dyadic" parts of the modes and fields should be common for all components. Consequently, proposed operator form of QFT offers possibility 
to impose condition for the electromagnetic field that leads to appropriate theoretical description of the experimental fact that no scalar or longitudinal 
photons have been observed. More concretely, instead of the well known Lorentz gauge $\partial _{\mu} \cdot \hat A^{\mu} = 0$, the condition one should take 
now reads: 
\begin{equation}
\hat k _{\mu , re} \cdot \hat a^{\mu} \cdot \vert \Psi \rangle = 0,
\end{equation}
where the quadri vector of operators $\hat k _{\mu , re} =({\hat s_{re} \over c_{re}} , {\hat {\vec k}} _{re})$. The annihilation operators $\hat a^{\mu}$ act 
non-trivially only in one of the four amplitude spaces that are attached to four components of the electromagnetic field, for others they are equal to $\hat I$, 
while the dyadic part is as in (6-7) ($\vert \Psi \rangle$ is the state of the electromagnetic field). For the particular choice of the coordinate system, 
adapted to the momentum of the mode $k _{\mu , re} = (k, 0, 0, k)$, one takes the four components of the electromagnetic field to be the so-called scalar, two 
transverse and longitudinal ones. Then, (in the standard notation of quantum theories where energy and time come after position and momentum) from the above 
condition it follows that:
$$
\hat a^{s} \vert n ^{s} \rangle \otimes \vert n ^{t 1} \rangle \otimes \vert n ^{t 2} \rangle 
\otimes \vert n ^{l} \rangle \otimes \vert \vec k _{re} \rangle \otimes \vert E_{\vec k _{re}} \rangle \otimes 
\vert \vec k _{im} \rangle \otimes \vert E_{\vec k _{im}} \rangle -
$$
\begin{equation}
- \vert n ^{s} \rangle \otimes \vert n ^{t 1} \rangle \otimes \vert n ^{t 2} \rangle \otimes \hat a^{l} \vert n ^{l} 
\rangle \otimes \vert \vec k _{re} \rangle \otimes \vert E_{\vec k _{re}} \rangle \otimes \vert \vec k _{im} \rangle \otimes 
\vert E_{\vec k _{im}} \rangle = 0 .
\end{equation}
Since the scalar and longitudinal annihilation operators act non-trivially in different Hilbert spaces, the last equation can be satisfied only if both 
$\vert n ^{s} \rangle$ and $\vert n ^{l} \rangle$ are the vacuum states, the meaning of which is that there are no longitudinal and scalar photons. 
Excitations of transverse components are not bounded by the considered gauge. 

If one wants to consider the quadri vectors of operators:
\begin{equation}
\hat k  _{{\mu}, re} = ( {\hat s \over c_{re}} , {\hat {\vec k}} _{re} ) ,
\end{equation}
and
\begin{equation}
\hat k _{{\mu}, im} = ( {\hat s \over c_{im}} , {\hat {\vec k}} _{im} ) ,
\end{equation}
and to do that for a whole field taken as one entity, then one should introduce new rigged Hilbert spaces in direct product within which the vectors of appropriate 
operators (total momenta and total energy) should be. In the case of multicomponent field each of them can have its own Hamiltonian and if there are many fields 
(with many components) in interaction, or there is mutual interference between real and imaginary parts, then the Hamiltonian becomes complicated. In general, the 
Hamiltonian can be the sum of many terms, all of which can depend on all involved operators, {\it e. g.} $ {\hat {\vec r}} _{re}$, ${\hat {\vec k}} _{re}$, 
${\hat {\vec r}} _{im}$, ${\hat {\vec k}} _{im}$ and $\hat t$, of all fields and their components. If the Hamiltonian depends on time, then the attached vectors 
$\vert E _{\vec k _{re}} \rangle$ and $\vert E _{\vec k _{im}} \rangle$ in $\vert t \rangle$ representation are no longer of the form 
$e^{{1\over {i \hbar}} E_{\vec k _{re}} \cdot t }$ and $e^{{1\over {i \hbar}} E_{\vec k _{im}} \cdot t }$. Then they are more general functions 
and this was discussed in [32]. Needle to say, the collision processes within the operator form of QFT, on the other side, one can represent by application of the 
creation and annihilation operators of involved fields.

Constraining equations might involve many partial derivations. Since the operators representing fields are diagonal in the dyadic part, partial derivatives 
$\partial _{\mu}$ should not be mapped to ${1\over i \hbar} [ \ \ \ , \hat k _{\mu , re} ]$ because if such commutators are applied on quantized field, the result 
would be equal to zero. Instead of these commutators, one should apply $ \hat k _{\mu , re}$ on fields from one side and, in order to find the solutions of the 
operator version of mentioned equations, take that all fields are self-adjoint operators. Finally, let us remark that there could be the case when the vectors 
$\vert \vec k _{re} \rangle $ (plain waves $e^{-{1\over {i \hbar}} \vec k _{re} \cdot \vec r _{re} }$) are no longer appropriate for 
the modes, {\it i.e.}, the modes are given by some other functions. 

\section{Conclusion}

We have proposed operator formulation of QFT that can be used for adequate description of quantum fields. All fields, including gravity with its metric field, 
as well, should be quantized according to the above given prescriptions and many problems that occur in standard formulation of QFT would be avoided. Regarding the 
comparison with standard treatment of classical fields, the state:
\begin{equation}
\int \int \vert n _{\vec k _{re} , \vec k _{im} } \rangle \otimes \vert \vec k _{re} \rangle \otimes \vert 
E_{\vec k _{re}} \rangle \otimes \vert \vec k _{im} \rangle \otimes \vert E_{\vec k_{im}} \rangle \cdot d\vec k _{re} \cdot d\vec k _{im} ,
\end{equation}
with $\vec k _{re} = \vec k$ and $\vec k _{im} = i \cdot \vec k$, corresponds to:
\begin{equation}
\int ( A_{{\vec k}} e^{-{1\over {i \hbar}} {\vec k} \cdot {\vec r} }  e^{+{1\over {i \hbar}} E_{{\vec k}} \cdot t } 
+ A^*_{{\vec k}} e^{+{1\over {i \hbar}} {\vec k} \cdot {\vec r}} e^{-{1\over {i \hbar}} E_{{\vec k}} \cdot t }) 
d{\vec k} .
\end{equation}
The "amplitude vector" $\vert n _{\vec k _{re} , i \cdot \vec k} \rangle$ of quantum field corresponds to the amplitude 
$A_{{\vec k}}$, and $A^* _{{\vec k}}$, of the classical field, the "mode vector" $ \vert \vec k \rangle $ corresponds to 
$e^{-{1\over {i \hbar}} \vec k \cdot \vec r }$ since $ \langle \vec r \vert \vec k _{re} \rangle = 
e^{-{1\over {i \hbar}} \vec k \cdot \vec r }$, the "mode vector" $ \vert i \cdot \vec k \rangle $ corresponds to 
$e^{+{1\over {i \hbar}} \vec k \cdot \vec r }$ since $ \langle i \cdot \vec r \vert i \cdot \vec k \rangle = 
e^{+{1\over {i \hbar}} \vec k \cdot \vec r }$ and the positive energy vector $\vert E_{\vec k } \rangle $ and negative energy vector 
$\vert E_{i \cdot \vec k } \rangle$ correspond to $e^{+{1\over {i \hbar}} E_{\vec k} \cdot t }$ and $e^{-{1\over {i \hbar}} E_{\vec k} \cdot t }$, 
respectively. 

Since we have proposed the formalism that treats real and imaginary quantities on an equal footing, let us stress that all our observables are of the form 
$\hat M _{re} \otimes \hat I _{im}$. On the other side, within this formalism different fields have their own set of the "spatio-temporal" Hilbert spaces in direct 
product, so the localization in a space and/or time point is represented by application of adequate projectors in the appropriate Hilbert spaces. In particular, if 
all involved fields should be localized at some moment $t_0$, then $\vert t_0 \rangle \langle t_0 \vert$ should be applied in all Hilbert spaces where the energy 
vectors appear.    
 
The building blocs of the proposed formalism are the mode operators, the spatio-temporal quadri vectors of operators of the real and imaginary coordinates and 
momenta and two non-commuting amplitude operators. With these, we believe, formal description of all situations concerning quantum fields can be constructed and 
if there is necessity to multiply the non-commuting operators, one should use ordered (symmetrized) product. This was discussed in [39].

\section{Acknowledgement}

The author acknowledges funding provided by the Institute of Physics Belgrade, through the grant by the Ministry of Education, Science and Technological Development 
of the Republic of Serbia.

\end{document}